%% file: tvcg_journal.tex
\documentclass[10pt,journal,compsoc]{IEEEtran}
\ifCLASSOPTIONcompsoc
  \usepackage[nocompress]{cite}
\else
  \usepackage{cite}
\fi
\ifCLASSINFOpdf
\else
\fi

  \pdfoutput=1\relax                   
  \usepackage{graphicx}                
  \DeclareGraphicsExtensions{.pdf,.png,.jpg,.jpeg} 

\graphicspath{{figures/}{images/}{./}} 

\usepackage[hidelinks]{hyperref}
\usepackage{amsmath}
\usepackage{amssymb}

\usepackage{wrapfig}

\usepackage[inline]{enumitem}
\usepackage{multirow}

\usepackage{ragged2e}

\input{content/debug}

\begin{document}
%
\title{Visual Explanations with Attributions and Counterfactuals on Time Series Classification}
%
%
%
%

\author{Udo Schlegel,
        Daniela Oelke,
        Daniel A.\ Keim, 
        and Mennatallah El-Assady
\IEEEcompsocitemizethanks{\IEEEcompsocthanksitem U.\ Schlegel, and D.\ A.\ Keim, University of Konstanz, Germany.\protect\\
E-mail: \{u.schlegel, daniel.keim\}@uni-konstanz.de.
\IEEEcompsocthanksitem D.\ Oelke, Applied Science University Offenburg, Germany.\protect\\
E-mail: daniela.oelke@hs-offenburg.de
\IEEEcompsocthanksitem M. \ El-Assady, ETH AI Center, Switzerland\protect\\
E-mail: menna.elassady@ai.ethz.ch}
\thanks{Manuscript December 01, 2022.}}

%
%

\markboth{Transactions on Visualization and Computer Graphics,~Vol.~XXX, No.~XXX, XXX~2022}%
{Schlegel \MakeLowercase{\textit{et al.}}: Visual Explanations with Attributions and Counterfactuals on Time Series Classification}

%



\IEEEtitleabstractindextext{%
\begin{abstract}
\input{content/00-abstract-pure}
\end{abstract}

}

\maketitle

\IEEEdisplaynontitleabstractindextext

%
\IEEEpeerreviewmaketitle

\IEEEraisesectionheading{\section{Introduction}\label{sec:introduction}}

\input{content/0-content}
\ifCLASSOPTIONcompsoc
  \section*{Acknowledgments}
\else
  \section*{Acknowledgment}
\fi

This work has been partially supported by the Federal Ministry of Education and Research (BMBF) in the VIKING (13N16242) project.

\ifCLASSOPTIONcaptionsoff
  \newpage
\fi



\bibliographystyle{IEEEtran}
\bibliography{library}
%



%

\begin{IEEEbiography}[{\includegraphics[width=1in,height=1.25in,clip,keepaspectratio]{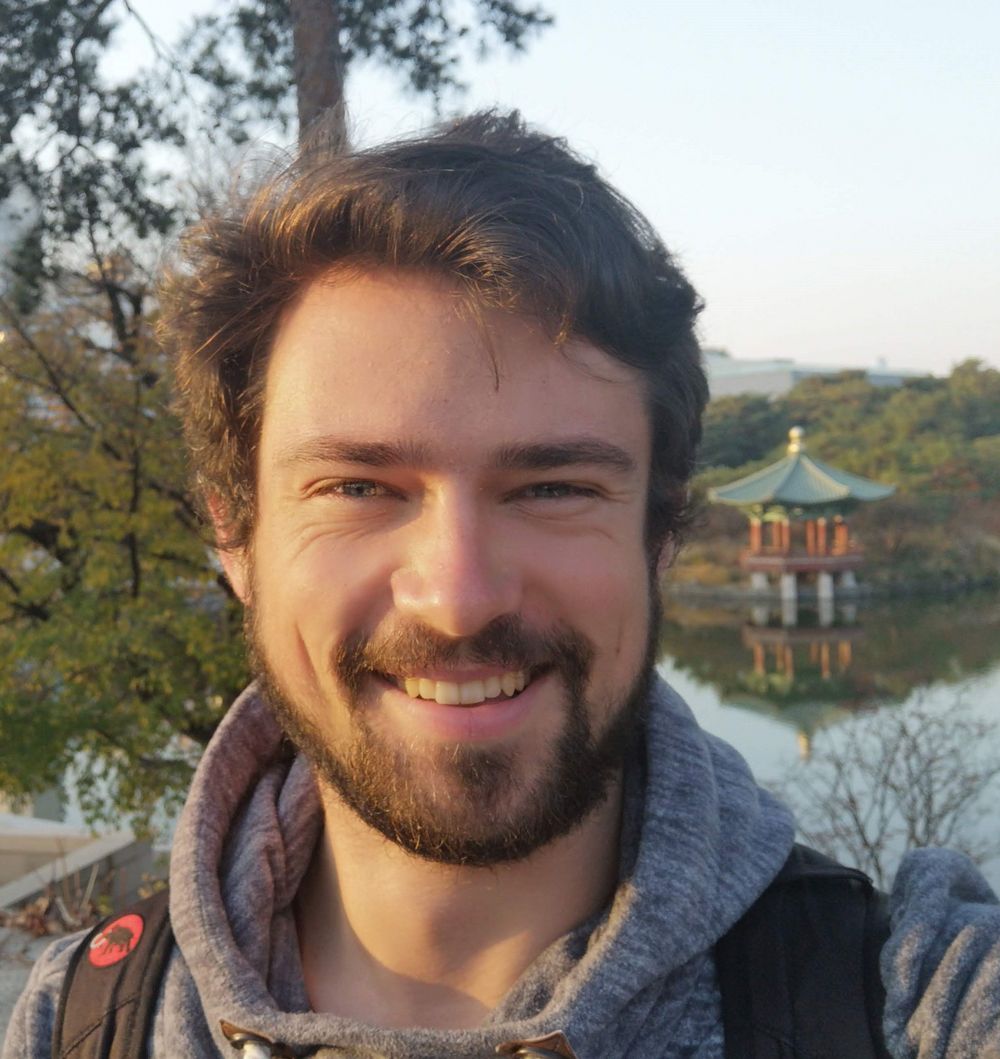}}]{Udo Schlegel} received his M.S. degree in Computer and Information Science from the University of Konstanz.
He is pursuing a Ph.D. in Computer Science at the University of Konstanz.
His research interest includes developing and evaluating explainable AI techniques for deep learning models on time series focusing on attribution approaches.
Further research interests include machine learning, visual analytics for deep learning, and deep learning applications.
\end{IEEEbiography}

\begin{IEEEbiography}[{\includegraphics[width=1in,height=1.25in,clip,keepaspectratio]{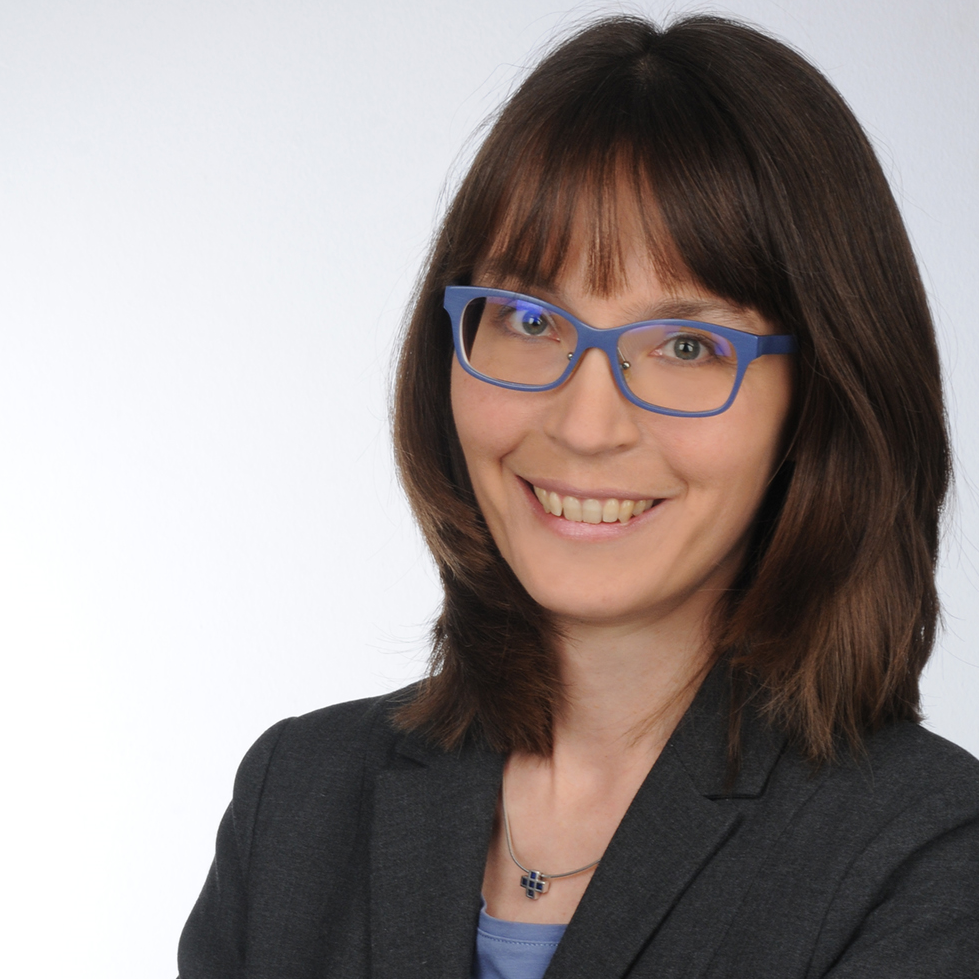}}]{Daniela Oelke}
Daniela Oelke is a professor for machine learning at Offenburg University of Applied Sciences. Her research interests lie in the areas of Explainable AI and Visual Analytics. She received her Ph.D. in computer science from the University of Konstanz. Before coming to Offenburg University, Dr. Oelke was employed at Siemens AG, as well as at the German Institute for International Educational Research (DIPF) in Frankfurt.
\end{IEEEbiography}

\begin{IEEEbiography}[{\includegraphics[width=1in,height=1.25in,clip,keepaspectratio]{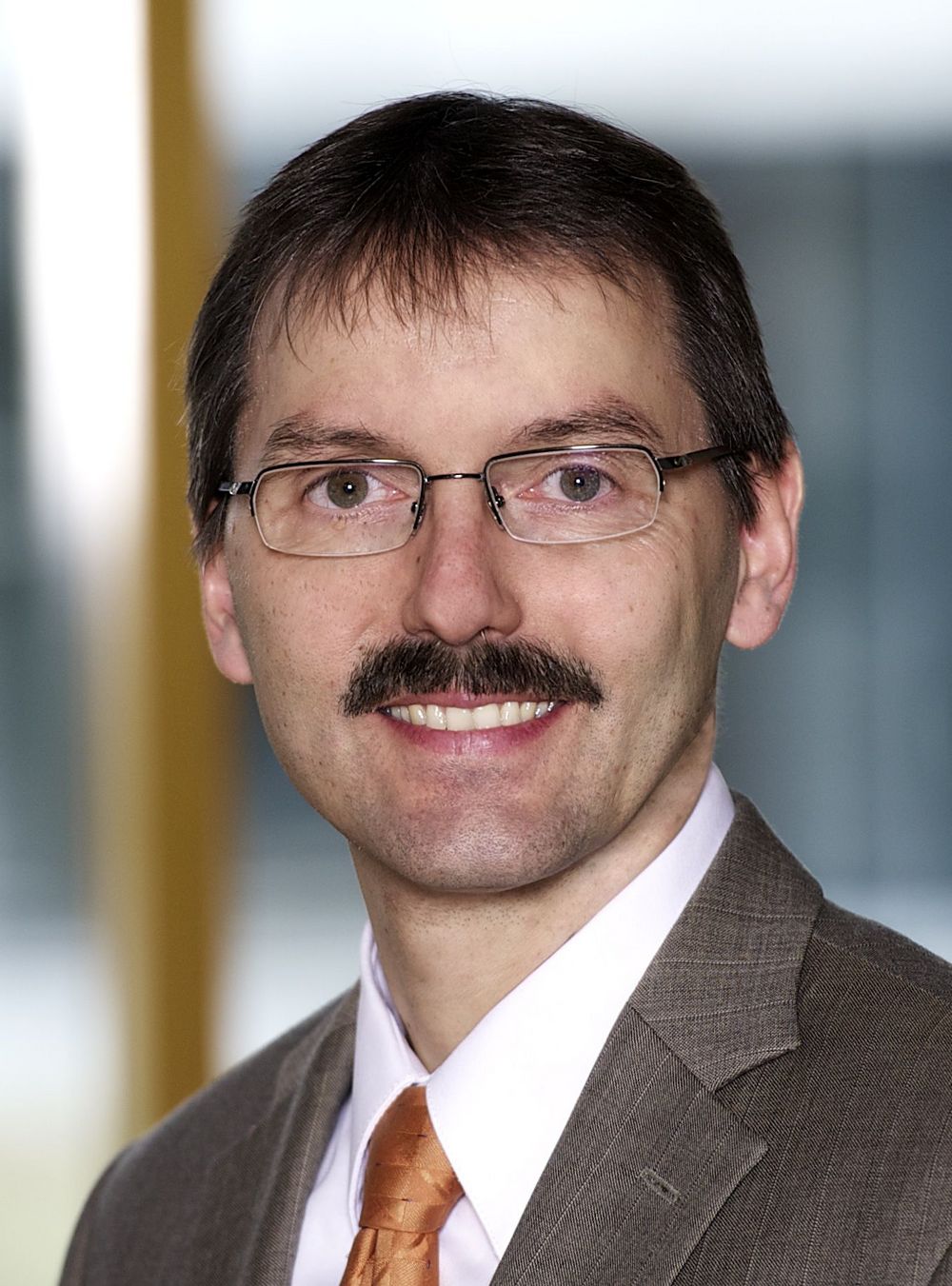}}]{Daniel A.\ Keim}
is a professor of the Data Analysis and Visualization Research Group at the University of Konstanz.
He has been actively involved in data analysis and information visualization research for over 25 years.
His services to the research community include papers co-chair of IEEE InfoVis 1999 and 2000, ACM SIDKDD 2002, IEEE VAST 2006 and 2019; general chair of InfoVis 2003; and associate editor of IEEE TVCG, IEEE TKDE, and Sage Information Visualization Journal.
Dr.\ Keim got his Ph.D.\ degree in Computer Science from the University of Munich, Germany.
Before joining the University of Konstanz, Dr.\ Keim was an associate professor at the University of Halle, Germany, and a Technology Consultant at AT\&T Shannon Research Labs, NJ, USA.
\end{IEEEbiography}

\begin{IEEEbiography}[{\includegraphics[width=1in,height=1.25in,clip,keepaspectratio]{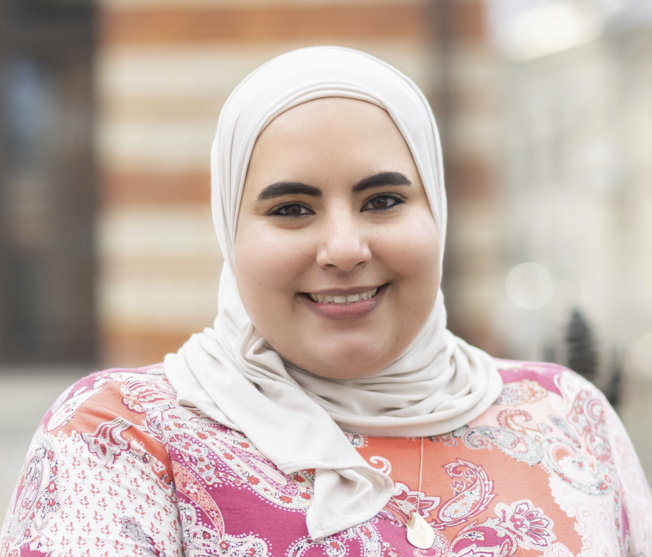}}]{Mennatallah El-Assady}
is a research fellow at the AI Center of ETH Zurich, Switzerland. Prior to that, she was a research associate and doctoral student in the group for Data Analysis and Visualization at the University of Konstanz, Germany, and in the Visualization for Information Analysis lab at the Ontario Tech University, Canada. She works at the intersection of data analysis, visualization, computational linguistics, and explainable artificial intelligence.
\end{IEEEbiography}




\end{document}

%% file: content/debug.tex
\usepackage{enumitem}
\usepackage{soul}
\usepackage[table]{xcolor}

\definecolor{c1}{rgb}{0,0.3,1}
\definecolor{c2}{rgb}{1,0,0.0}
\definecolor{c3}{rgb}{0.16, 0.5, 0.0}
\definecolor{c4}{rgb}{0.5, 0.5, 1}

\def\clr#1#{\@clr{#1}}
\def\@clr#1#2{~{\fboxsep0mm\fbox{\colorbox#1{#2}{\phantom{X}}}}}

\def\cssclr#1{{\fboxsep0mm\fbox{\colorbox[HTML]{#1}{\phantom{X}}}}}

%% file: content/00-abstract-pure.tex
\justifying 
With the rising necessity of explainable artificial intelligence (XAI), we see an increase in task-dependent XAI methods on varying abstraction levels. 
XAI techniques on a global level explain model behavior and on a local level explain sample predictions.
We propose a visual analytics workflow to support seamless transitions between global and local explanations, focusing on attributions and counterfactuals on time series classification.
In particular, we adapt local XAI techniques (attributions) that are developed for traditional datasets (images, text) to analyze time series classification, a data type that is typically less intelligible to humans.
To generate a global overview, we apply local attribution methods to the data, creating explanations for the whole dataset. 
These explanations are projected onto two dimensions, depicting model behavior trends, strategies, and decision boundaries.
To further inspect the model decision-making as well as potential data errors, a what-if analysis facilitates hypothesis generation and verification on both the global and local levels.
We constantly collected and incorporated expert user feedback, as well as insights based on their domain knowledge, resulting in a tailored analysis workflow and system that tightly integrates time series transformations into explanations.
Lastly, we present three use cases, verifying that our technique enables users to (1)~explore data transformations and feature relevance, (2)~identify model behavior and decision boundaries, as well as, (3)~the reason for misclassifications.

%% file: content/0-content.tex
\input{content/1-introduction.tex}
\input{content/2-related-work.tex}
\input{content/3-background.tex}

\input{content/4-1-technique}

\input{content/4-2-demonstrator}

\input{content/5-evaluation.tex}
\input{content/6-discussion.tex}
\input{content/7-conclusion.tex}

%% file: content/1-introduction.tex

\IEEEPARstart{D}{eep learning} (DL) achieves state-of-the-art performance in various tasks and domains such as computer vision (e.g., autonomous driving~\cite{huval_empirical_2015}) and natural language processing (e.g., machine translation~\cite{luong_deep_2015}).
Due to such cutting-edge applications, more and more fields incorporate deep learning models, e.g., time series forecasting (predictive maintenance~\cite{mobley_introduction_2002}).
However, time series are often \textit{not intuitively intelligible}, and thus manually debugging such models is a tedious task.
In many cases, transformations such as Fourier~\cite{bracewell_fourier_1986} or SAX~\cite{lin_symbolic_2003} are applied to the data to convert them into a more human-understandable abstraction.
These abstractions help to identify certain characteristics (e.g., frequencies), yet transformations cannot guarantee insights. 
Mainly, real-world sensor data can consist of various overlapping properties like periodicity.  
Class characteristics based on these single time series features are typically challenging to distinguish and are not guaranteed to be used by a classifier.
Thus, domain knowledge is regularly very focused on specific properties that are useful for the task at hand.
However, such domain knowledge is only sometimes readily available and is not guaranteed to, e.g., make classes distinguishable.
Thus, complex algorithms, like neural networks, need an in-depth inspection of their decision makings to provide possible insights into the data.


DARPA introduced the explainable artificial intelligence (XAI) initiative~\cite{gunning_explainable_2016} to foster and accelerate research around the topic of explainable machine learning (ML) as well as increase trust in AI models.
Thus, such initiatives foster gaining an understandable explanation of the model on the one hand and more insight into abstract data on the other hand.
Such explanations can be generated on a global and local level~\cite{guidotti_survey_2018}.
The global level introduces methods to describe the overall decision-making of the model~\cite{guidotti_survey_2018}.
The local level, conversely, explains only single decisions, e.g., the prediction of a single data sample~\cite{guidotti_survey_2018}.
An approach for global explanations using visual analytics is RuleMatrix~\cite{ming_rulematrix_2019}.
The decision-making can be inspected by extracting basic decision rules from a classifier.
However, the current approach is time-consuming and works best on tabular data with limited features.
In most cases, either a global or local explanation is provided~\cite{hohman_summit_2019, strobelt_seq2seq_2019}, and a switch between both is often tedious.
Thus, enabling interaction between global and local explanations supports explaining model decisions while still allowing for an in-depth analysis of the data.


We propose a system based on a workflow to seamlessly integrate the transition between global and local explanations for time series, incorporating a technique to generate global explanations based on local attributions.
As a baseline for our explanations, we include state-of-the-art attribution techniques for images, text, and tabular data, such as LRP~\cite{bach_pixel_2015} and SHAP~\cite{lundberg_unified_2017}, as there are only a limited amount of time series techniques.
However, our technique can provide a global explanation based on any attribution method.
In particular, we propose an approach designed to generate global explanations for time series classifiers.
Our technique uses projection methods, e.g., PCA, to generate two-dimensional visualizations based on local explanations.
The workflow further presents a way to explain a time series model interactively on the global level as an overview for applied data and, on a sample, providing local level explanations.  
Global explanation projections facilitate getting an overview and finding decision boundaries for models.
For instance, through the interactive exploration of regions in these projections on the local level, decision borders can be found by generating counterfactuals~\cite{byrne_counterfactuals_2019} changing the time series sample.
Our approach combines local XAI techniques and projections to generate global explanations to represent and test the model representation of the data.
Further, by selecting interesting samples (e.g., overlapping classes) at the global level, the local explanation of these can be inspected and compared.
At the local level, the sample can be modified by the user based, e.g., on the attribution scores.
Such a modified sample can then be projected again in the global explanation to verify or reject the hypothesis about the decision boundaries of the model.
Through such interactions, a hypothesis can first be generated, verified, or rejected (what-if in local and back projection in global).
Our presented use cases show the technique's applicability on time series, indicating more comprehensive explanations for, e.g., misclassifications.

We contribute:
\begin{itemize*}[nosep]
    \item[(1)] A visual analytics workflow to bridge global and local explanations for time series. 
    \item[(2)] An overview technique based on global explanations produced with local \textit{attributions}.
    \item[(3)]  A local \textit{what-if} analysis to automatically and interactively generate \textit{counterfactuals} for time series.
\end{itemize*}
A time series application of our proposed workflow is presented with use cases that can be explored \href{https://visual-explanations.time-series-xai.dbvis.de/}{online}\footnote{Demo: \url{https://visual-explanations.time-series-xai.dbvis.de}}.


%% file: content/2-related-work.tex
\section{Related Work}
Our related work generally introduces important XAI concepts and presents the combination of XAI and VA.
Most XAI approaches and VA techniques target intelligible data, e.g., images or text, neglecting time series.

\textbf{Explainable Artificial Intelligence --}
In machine learning, the definition of explainability is often ambiguous and, in most cases, includes drastic definition differences~\cite{lipton_mythos_2018}. 
However, most definitions try to incorporate specific motivations such as fairness, privacy, reliability, and trust building~\cite{doshivelez_towards_2017} and thus follow the same larger goals to ensure these aspects~\cite{mittelstadt_explaining_2019}.
XAI methods consist of various techniques applied ad-hoc or post-hoc onto black-box models~\cite{adadi_peeking_2018}.
In recent years, the number of post-hoc XAI methods increased with \textit{explainers} aiming to achieve explainability for simple tasks~\cite{spinner_explainer_2019}.
Such methods have become increasingly popular due to the upsurge of deep neural networks and the ability of XAI methods to be applied to trained models~\cite{spinner_explainer_2019}.
Guidotti et al.~\cite{guidotti_survey_2018} categorize the methods into eight groups (e.g., feature importance, saliency masks) with various properties. 
One of the proposed properties describes the XAI method's level, either on the global model or on local samples, to explain the decision-making~\cite{guidotti_survey_2018}.
For example, many XAI methods highlight feature attribution for single samples, such as LIME~\cite{ribeiro_why_2016}, which offers a limited view of the model as it only shows a local explanation based on one sample.
Other methods on images present saliency masks~\cite{simonyan_deep_2013}, which display the feature importance of the attention on an image as a heatmap~\cite{bach_pixel_2015}.
However, such local explanations are generally not robust and can be fooled~\cite{subramanya_fooling_2019} or exploited~\cite{slack_how_2019}.

Furthermore, only limited work exists on the automatic evaluation of explanations~\cite{mohseni_survey_2018}.
%
In most cases, such evaluations are often conducted by perturbating test data with the corresponding explanations to inspect the performance change in computer vision~\cite{vu_evaluating_2019}, text data~\cite{samek_evaluating_2017}, and time series~\cite{schlegel_towards_2019}.
Ribiero et al.~\cite{ribeiro_anchors_2018}, however, discuss why such explanations and their perturbation evaluation are insufficient to gain insights into the model's internals.
Thus, an automated evaluation approach can be considered as a first step to highlight techniques working better than others on the model and data. 
However, humans need further inspection to be useful~\cite{mohseni_survey_2018}.

\textbf{Visual Analytics for XAI --}
Visual analytics includes the human at various steps of the data analysis and machine learning workflow~\cite{keim_visual_2008}.
Through various interactions, users can explore data, generate and test a hypothesis, and extract knowledge~\cite{sacha_knowledge_2014}.
Furthermore, humans can work not only with the data but also with models trained on the data for understanding, diagnosis, and refinement~\cite{liu_towards_2017}.
Addressing these tasks, Spinner et al.~\cite{spinner_explainer_2019} introduce a framework and pipeline to steer the explanation process of XAI and support the tasks' understanding, diagnosis, and refinement.
However, the framework lacks an extension towards incorporating the interaction between global and local explanations to integrate the human into the analysis loop for hypothesis generation and testing like Sacha et al.~\cite{sacha_knowledge_2014}.

VA has various approaches explaining the decision-making of models or steering them into a user-preferred direction, such as Protosteer~\cite{ming_protosteer_2019}.
Especially, Liu et al.~\cite{liu_towards_2017} introduce the three tasks of understanding, debugging, and refinement to guide the process of an ML model in VA.
Further, Hohman et al.~\cite{hohman_visual_2018} survey VA works that focus on an interactive analysis of deep learning models.
Their survey is based on an interrogative technique to categorize works, e.g., if these support an understanding or decision explanation.
Further, related approaches surveyed in the previous work often operate only on local explanations.
A related approach for sequence data, mostly text, is LSTMVis~\cite{strobelt_lstmvis_2018} as it provides activations of LSTM cells throughout sequence inputs.
Another approach for sequence data provides Seq2Seq-Vis~\cite{strobelt_seq2seq_2019} and focuses on machine translation by visualizing other translation possibilities of the model as well as paths through the explanation space.
An approach for images presents Summit~\cite{hohman_summit_2019} that combines feature visualizations of convolutional neural networks for global explanations with sample activation for local explanations.
RuleMatrix~\cite{ming_rulematrix_2019} presents a global explanations solution for tabular data by extracting decision rules from a trained model to visualize feature importance, attributions, and distributions.
In contrast, scaling and explaining decision rules for time series is rather complicated.
For time series, ProtoSteer~\cite{ming_protosteer_2019} proposes an interactive analysis to enrich models with user-steered prototypes for partially explainable components and, in exchange, for performance.
Collaris et al.~\cite{collaris_strategyatlas_2022} introduce StrategyAtlas, which uses projected feature importance values through UMAP~\cite{mcinnes_umap_2018} to highlight model strategies.
Such strategies demonstrate the model's internal clustering of the data.
We use a similar approach. 
However, due to possible artifacts of UMAP due to incorrect parameters, we cover various techniques to mitigate such issues.
Further, our work focuses primarily on the transition between local and global explanations to facilitate the analysis of models and data on both levels.

\textbf{Visual Analytics for Time Series --}
Basic line plots are generally the go-to solution for time series.
However, such visualizations are challenging to scale for long time series or many time series samples.
An approach to overcome these challenges uses SAX Navigator~\cite{ruta_sax_2019} by applying SAX~\cite{lin_symbolic_2003} to transform time series into an abstract symbolic representation and cluster the results to explore the data on various levels.
Other proposals use dimension reduction techniques to visualize time series as paths such as Time Curves~\cite{bach_time_2015} to cope with long time series and hard-to-read, as well as compare line plots.
Such time paths can be used for various tasks such as time series clustering~\cite{ali_timecluster_2019}.
Also, such a technique can be applied to visualize decisions of algorithms to comprehend choices and outcomes~\cite{hinterreiter_exploring_2020}.
One major drawback of time paths is the clutter produced by many overlapping time paths.
Our approach focuses on many time series samples and, thus, on dimensionally reducing the time series to a single point.

\textbf{XAI for Time Series --}
Theissler et al.~\cite{theissler_explainable_2022} provide a comprehensive state-of-the-art survey and review of possible XAI methods for time series, including a categorization towards the level of explanation.
We consider this categorization as the baseline for choosing possible techniques and focus on the ones for attributions and counterfactuals.
Thus, we extract and collect the possible methods for time-point-based and instance-based explanations from Theissler et al.~\cite{theissler_explainable_2022}.
We mitigate subsequence-based explanations as these often need specific architectures.
For further information on the topic, we recommend the survey~\cite{theissler_explainable_2022}.

\textbf{Combined XAI and VA for Time Series --}
The combination of visualizations, visual analytics, and explainable AI for time series got a boost in the last few years.
Schlegel et al.~\cite{schlegel_time_2021} discusses various approaches of XAI for time series and explores the recent visualization techniques for attributions on time series.
Especially, Assaf et al.~\cite{assaf_mtex_2019}, and Schlegel et al.~\cite{schlegel_towards_2019} propose early options on how to visualize attributions for time series data.
However, most of these approaches are rather tricky to interpret for most users, experts, and non-experts. 
Attribution visualizations without further interaction options are limited in their communication of model behavior~\cite{schlegel_time_2021} as often the attribution techniques heavily influence the attributions~\cite{schlegel_towards_2020}.
Siddiqui et al.~\cite{siddiqui_tsviz_2019} propose with TSViz a technique to investigate CNNs applied to time series.
Through attributions on the input data based on either the output or inner filters, they visualize the most relevant parts of the time series on the line chart.
However, their approach focuses heavily on CNNs and needs some tweaking for other methods.
We seamlessly support all kinds of neural network architectures.

%% file: content/3-background.tex
\section{Background And Characterization}
The classification of \textit{non-intelligible} data (e.g., time series) using complex models encompasses problems regarding the underlying data, the applied model, and the user tasks.
For instance, time series data often contains noise, sensor errors, and arbitrary segmentations producing even more errors~\cite{mobley_introduction_2002}.
Thus, raw time series are often complicated to interpret due to influencing error factors.
Data processing and transformations applied by experts often lead to the first insights into the data.
However, transformations such as Fourier-Transformation~\cite{bracewell_fourier_1986} often need constraints, e.g., segmentation, which is not readily applicable for critical real-world applications.
Therefore, fast models need to be trained on the raw data in many use cases.
For instance, failure detection on top of machines is often trained on raw sensor data, resulting in complex, tailored models.
Especially, neural network models such as LSTM FCN~\cite{karim_lstm_2017} achieve state-of-the-art performance with an acceptable prediction speed.
Such models, though, are generally not interpretable by design.
Not only is the data hard to interpret, but also models lack comprehensible decisions.

Based on these findings, we discuss the needs of users who apply time series classification models daily and aim to facilitate their work as much as possible.

\subsection{Users, Data, and Model}

\textbf{Users --} 
We focus on the two types of users, domain experts (DE) and data scientists (DS), working with time series.
Domain experts are users applying pre-trained models on their daily work for, e.g., predictive maintenance~\cite{mobley_introduction_2002}.
Data scientists cover a range of users, including developers building models on raw time series data.
Other users, such as end-users, shift the focus from debugging and refining models toward understanding.
This understanding is often provided by domain- and task-specific visualizations, e.g., Schlegel et al.~\cite{schlegel_towards_2020}, to target the specific user group.
Our approach primarily focuses on the general applicability of debugging and refinement of data and models.
Domain knowledge supports the understanding as we mainly look at model and data connections. 
End-users are, therefore, only addressed to a limited extent.
During the development of the system, we were in contact with DE and DS to steer the research direction.

\textbf{Data --} 
Sensor data is the base for many time series applications and is available as open-access datasets.
In many real-world applications, interviewed users often reported a need for labels for such time series data and labeled their data by themselves on only a limited amount of samples.
However, even with labels, many models run into unknown failures due to previously mentioned errors.
As time series are \textit{non-intelligible}, we target such data to support the analysis of models applied on raw time series with limited domain knowledge to enhance workability with such data.
We further focus on time series classification to support the supervised learning of segmented time series data with labels to enable understanding and debugging.
Our approach focuses mainly on uni-variate time series to present the overall methodology.
However, the proposed approach can also be extended to multi-variate data.

\textbf{Model --} 
Our approach focuses on models applied to raw time series data to avoid as many transformations as possible.
Additionally, we target deep neural networks as these provide state-of-the-art performance on many time series classification and forecasting tasks, e.g., LSTM-FCN~\cite{karim_lstm_2017}.
However, we do not further constraint the architecture as transformer networks~\cite{vaswani_attention_2017}, convolution neural networks, and recurrent neural networks present nearly state-of-the-art results on time series in many cases~\cite{lai_modeling_2018}.

\subsection{User Needs}

Based on previous domain expert interviews and experience, we emphasize three needs (N) users have while working on their analysis.

The \textbf{\textit{needs}} we came accross:
\textbf{\noindent(N1) \textit{Explore} data transformations and the internal model representations of data.}
Time series are inherently \textit{non-intelligible} and suffer from heavily needed domain knowledge or transformations changing the data domain.
Fourier transformations are the most commonly used way to process the data to a more accessible, readable domain.
However, even such transformations can take time to interpret.
Thus, the first need of users we identified is to comprehensively explore the data, transformations, and the corresponding internal model representations of the data.

\textbf{\noindent(N2) \textit{Identify} the model behavior on known and unknown data.}
As state-of-the-art models get more complex during the last few years and their inner workings often need to be more easily understandable, another need is the fast identification of the model behavior for the applied data.
Thus, users require an overview of the decisions and internals of a model during application on data to get insights into the models' behavior to be able to understand, e.g., misclassifications.

\textbf{\noindent(N3) \textit{Understand} sample misclassifications to investigate into the model.}
After identification of model behavior, users need to be able to investigate the decisions of models in-depth on samples to understand, e.g., misclassifications.
Thus, like in the Shneiderman mantra~\cite{shneiderman_eyes_2003}, the overview helps to guide to the details a user requires to understand the models' predictions overall.
Afterward, information about the single sample prediction is essential to support users in understanding the models' local decisions.

%% file: content/4-1-technique.tex
\begin{figure*}[tb]
 \centering 
 \includegraphics[width=\linewidth,trim={0.7cm 3.6cm 0.7cm 5.4cm},clip]{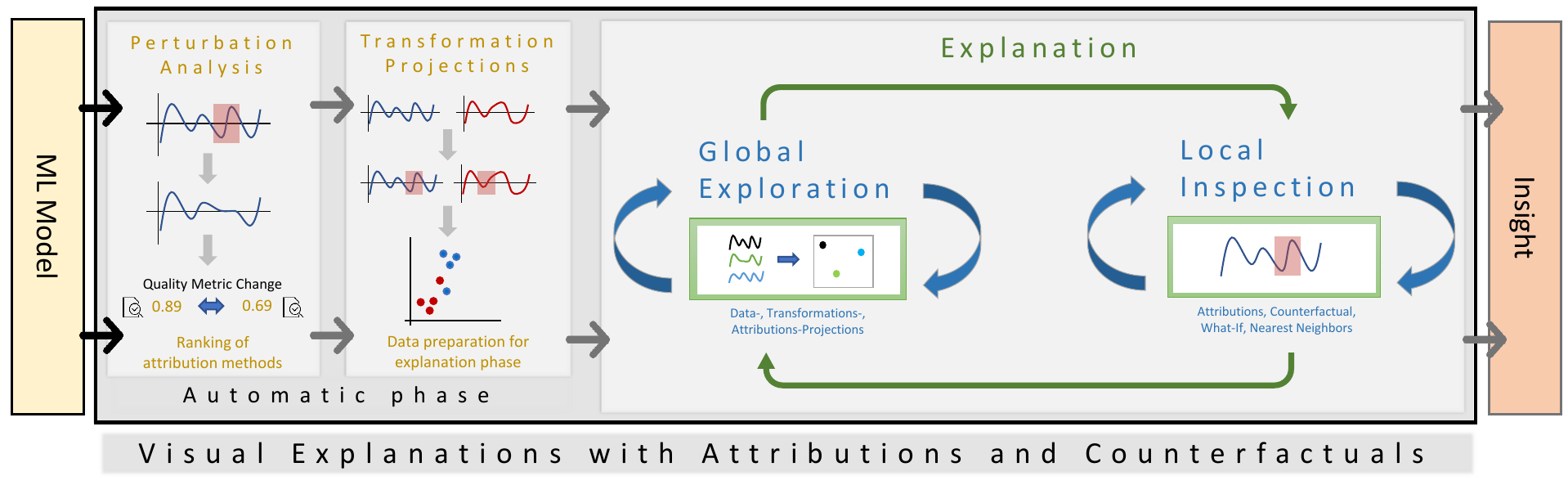}
 \vspace{-2em}
 \caption{Workflow for Visual Explanations with Attributions and Counterfactuals: starting with an ML model,
 the workflow automatically ranks attributions methods using a perturbation analysis to get comparable scores, then projects the attributions and transformations into a two-dimensional space. In the manual analysis, the explanation loop starts with two subloops on the global exploration and local inspection level to ensure the generation of explanations.}
 \vspace{-1em}
 \label{fig:workflow}
\end{figure*}

\section{Visual Explanations with\newline Attributions and Counterfactuals}
We propose a workflow to conceptualize global exploration and local inspection interactions to enable analysts to examine their time series models.
As the global explanation level is often difficult to obtain for complex models, we apply projections to create global explanations based on local attributions similar to StrategyAtlas~\cite{collaris_strategyatlas_2022}.

\textbf{Workflow --}
Our workflow grounds in the KDD process~\cite{piateski_knowledge_1991}, the explAIner pipeline by Spinner et al.~\cite{spinner_explainer_2019} and extends these with techniques used by experts, including considerations of our previous work~\cite{schlegel_time_2021}.
We extend the XAI component for global and local interactions on top of the data, the model, and the output to explore counterfactuals, attributions, and predictions.
Our proposed workflow consists of three stages, as seen in \autoref{fig:workflow}, in which the first two stages form the automatic phase, while the user purely drives the last stage.
The automatic phase consists of the perturbation analysis and transformation projection stages.
These two process the data and extract information for the last stage.
At first, the perturbation analysis (automatic evaluation) of local attribution methods generates attributions for the dataset and pre-ranks suiting techniques based on their evaluation scoring for further analysis.
Next, the transformation projection stage applies various projection techniques on the raw time series, the Fourier-transformed data, and the attributions to reduce the dimensionality to two.
After the automatic phase, the explanation phase incorporates the user into the explanation process~\cite{elassady_towards_2019}.
In the first global exploration, the previously calculated results visualize an overview of the data, the transformations, and the attributions.
Through the global exploration or by domain knowledge about the data, the user gets to the local inspection in which the in-detail analysis can be done using, e.g., a what-if analysis similar to the What-If tool~\cite{wexler_what_2019}.
The following sections describe the components of the workflow in more detail.




\subsection{Automatic Phase}

The automatic phase is pipelined into two stages.
In the first stage, various attribution techniques extract attributions as explanations for the model, and the perturbation analysis calculates comparable scores of these attributions.
Then, in the second stage, the generated attributions get projected into two-dimensional space for the global stage using various dimension reduction techniques.

\textbf{Perturbation Analysis --}
At first, all selected attribution 
\begin{wrapfigure}[9]{r}{2.5cm}
\includegraphics{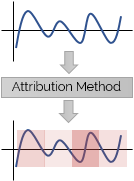}
\end{wrapfigure}
methods generate attributions for the model based on the selected data.
The extracted attributions can be generated in various ways but always attribute a single relevance score to every input value based on the output of the model~\cite{theissler_explainable_2022}.
E.g. for an input time series sample of length 500, an attribution contains 500 relevance scores.
For more information, Theissler et al.~\cite{theissler_explainable_2022} survey and explain various techniques which are applicable to time series.

A perturbation analysis is a fundamental approach to evaluate such attributions on their correctness towards a model's behavior with a score~\cite{samek_evaluating_2017}.
In general, a perturbation modifies (remove, increase, decrease) the most relevant features based on the attributions from a sample to change the prediction of a model~\cite{vu_evaluating_2019}.
The change in performance (quality metric score) highlights if the feature was relevant for the prediction~\cite{samek_evaluating_2017}.
Such a perturbation analysis can be done fully automatically with just a few modifiable parameters~\cite{schlegel_towards_2019}.
Especially, fully automatic procedures enable to compare the performance of various attribution methods~\cite{samek_evaluating_2017}.
For instance, significant changes in the quality metric score (e.g., accuracy) highlight techniques that accurately describe the model's decision-making~\cite {vu_evaluating_2019}.
Random importance perturbation can be used to get a baseline to compare the methods against~\cite{schlegel_towards_2019}.
These changes help to sort the attribution techniques and inspect which performs best and if good-performing methods are better than random measures.
However, as these attribution methods only work on the data in collaboration with the model, the model's inner workings are not explicitly explained.
Further, attributions are sometimes hard to interpret visually, for instance, if using time series as time points are often shown as important without the context~\cite{schlegel_time_2021}.
Nevertheless, through such a perturbation analysis, attribution methods can be evaluated and shown to be trustworthy explanations~\cite{theissler_explainable_2022}.

Such perturbation analyses have been done in computer vision~\cite{vu_evaluating_2019}, text data~\cite{samek_evaluating_2017}, and time series~\cite{schlegel_towards_2019}.
Schlegel et al.~\cite{schlegel_towards_2019} propose different perturbation approaches categorized into two groups (point and time) to use for time series to evaluate attribution methods on time properties.
As stated before, the general idea for a perturbation is a change of data values of a sample to something else.
\begin{wrapfigure}[8]{r}{2.5cm}
\includegraphics{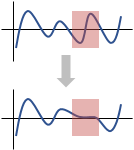}
\end{wrapfigure}
For point perturbation, relevant time points (single data values) are perturbed to, e.g., zero.
In literature, e.g., in computer vision~\cite{hooker_benchmark_2019}, the value needs to be changed to a non-information holding value.
However, in time series, such a value is hard to find, as data does not always have hard possible value boundaries.
Schlegel et al.~\cite{schlegel_towards_2020} also propose to modify them to their inverse, max, or min to have a larger spectrum of possible changes.
To incorporate also the time component of time series, Schlegel et al.~\cite{schlegel_towards_2019} propose the time perturbation with another approach to perturb the data values.
For more information on such perturbation analysis, we recommend Schlegel et al.~\cite{schlegel_towards_2020} and Theissler et al.~\cite{theissler_explainable_2022}.

Such an evaluation is vital to ensure the trustworthiness of the attribution methods at hand~\cite{samek_evaluating_2017}.
For instance, a faulty explanation misleads users into wrong insights of the model and worsens the understanding~\cite{mittelstadt_explaining_2019}.
Robust explanations are a preferable entry point for further analysis and essential to be able to understand the models behaviour~\cite{mittelstadt_explaining_2019}.


\textbf{Transformation Projections --}
In the next stage, the data is used to extract more transformations together with the attributions, e.g., the Fourier transformation.
Also, the models' activations of, e.g., the last layer before the softmax can be extracted and used as transformation.
These transformations, attributions, and the raw data are projected using various dimension reduction techniques to two dimensions.

\begin{figure}[hb!]
 \centering 
 \includegraphics[width=\linewidth]{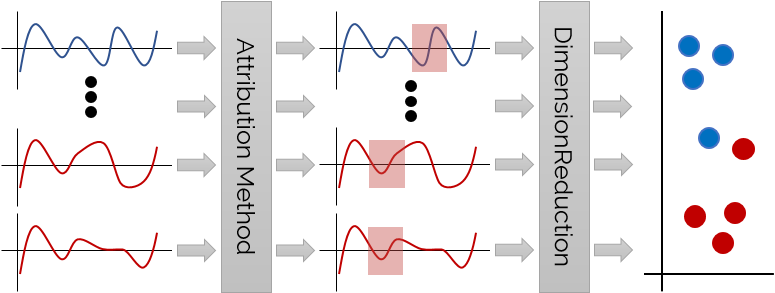}
 \caption{Generating global explanations for time series; first, an attribution method like LIME~\cite{ribeiro_why_2016} generates feature attributions for each sample in the whole dataset; next, all calculated attributions are projected using a dimension reduction technique like UMAP~\cite{mcinnes_umap_2018}; a scatter plot then demonstrates an inner representation of the data for the model.}
 \label{fig:technique}
\end{figure}

\autoref{fig:technique} shows the general concept of the technique.
The attribution technique takes all samples from the data and calculates the attributions for every sample.
This new attribution data is then projected on two dimensions to visualize these in a global overview of the data.
Depending on the dimension reduction technique, various properties like neighborhood preservation or distances hold and can help to mitigate artifacts in the projection as clusters are shared over different techniques showing cluster properties.

\subsection{Global and Local Explanations}
After a successful automatic phase, the explanation loop starts with the global exploration of the projections.
We first introduce the global exploration as an overview method based on projections of local attributions.
Afterward, we present the local inspection to show the possibilities and limits of local attribution methods

\begin{figure*}[h!t]
 \centering 
 \includegraphics[width=\linewidth,trim={0cm 3.4cm 0cm 0cm},clip]{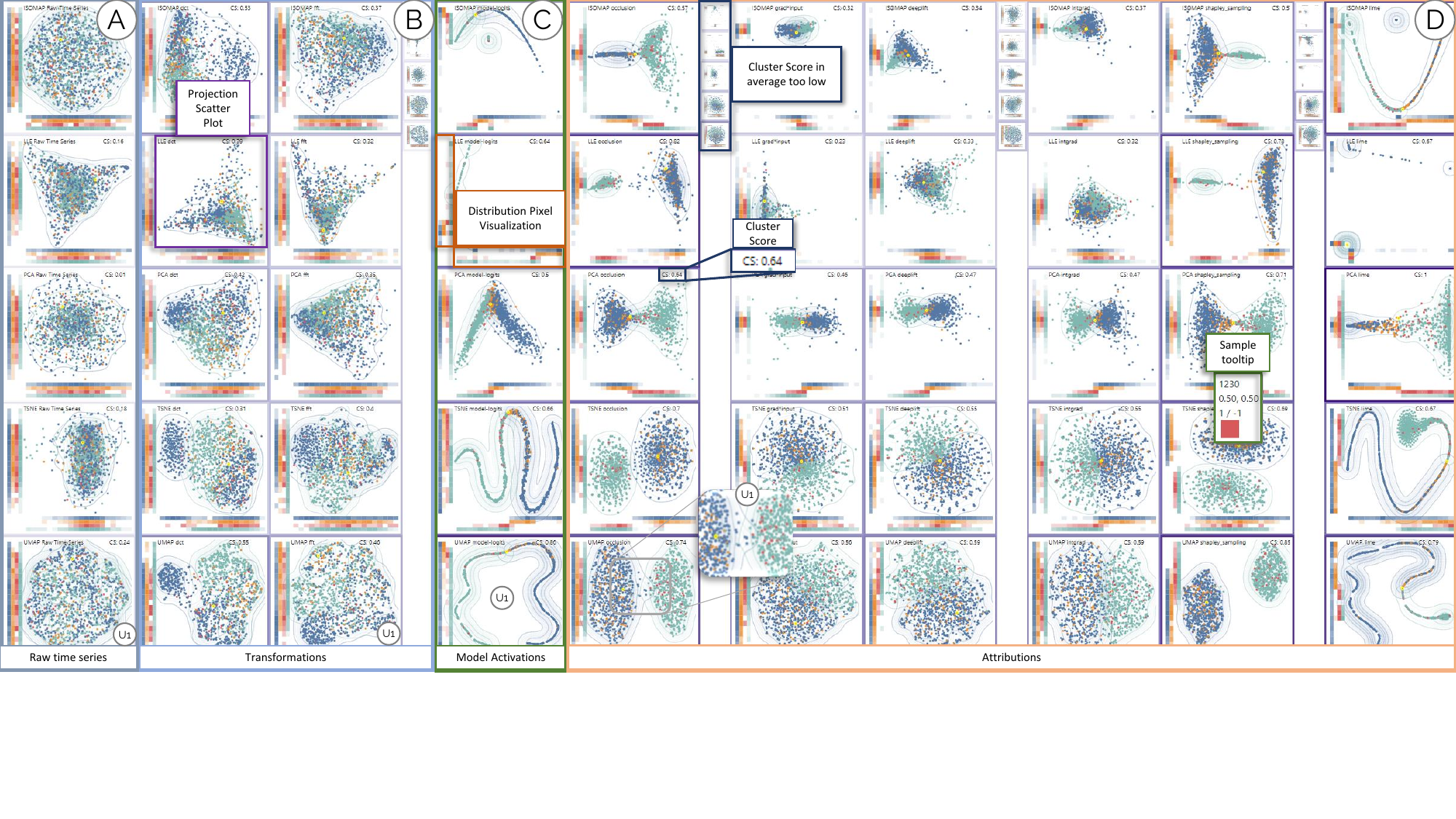}
  \vspace{-2em}
 \caption{Data (A), transformations (B), model activations, and attributions (D) projections as global explanation of a model for the FordA dataset. The color of the points corresponds to the confusion matrix result of the classifier. \cssclr{4e79a7} as correct normal state (true positive); \cssclr{76b7b2} as correct abnormal state (true negative); \cssclr{f28e2c} as incorrect normal state (false positive); and \cssclr{e15759} as incorrect abnormal state (false negative).}
 \label{fig:global-explanation}
  \vspace{-1em}
\end{figure*}

\textbf{Global Exploration --}
Global explanations enable to understand models in total, e.g., by visualizing decision boundaries or by comprehensible decision rules~\cite{guidotti_survey_2018}.
Decision rules are, in most cases, favorable~\cite{ming_rulematrix_2019}; however, the extraction process has a very high computation time and is often also hard to interpret for time series.
For most data, decision boundaries are simpler to comprehend and show shortcomings of models (critical samples near borders).
However, decision boundaries for high-dimensional or time series data are often hard to visualize.
Such boundary visualizations have to solve the same challenges as the high-dimensional data and need further abstractions to display the decision borders in the same visualization.

Only a few techniques for complex models propose global explanations and solutions to overcome such visualization challenges.
One of these examples is ANCHORS~\cite{ribeiro_anchors_2018}, which enhances LIME~\cite{ribeiro_why_2016} by searching for decision boundaries for certain classes so-called anchors.
However, the extracted rules consist of decision rules and, thus, are of rather limited use for e.g., time series.
For a promising explanation, these explanations rules need further transformations, which introduce other problems such as limited interpretability or loss of information.

We incorporate attributions of a model on data to generate a global overview of the models' strategies similar to StrategyAtlas~\cite{collaris_strategyatlas_2022}.
We take the results of the automatic phase, the pre-ranking, and the projections to visualize the data, transformations, and attributions.
Due to, for example, artifacts or incorrect parameters, we propose to use not only one dimension reduction technique but a wider variety of approaches.
By not only using a manifold, we also get more robust projections and options for users to dig into different projections and their results.
Thus, outliers and clusters can be more easily investigated in the two-dimensional space.
Further, comparing the different results of the techniques enables further investigation into the results.
For example, an outlier in the attribution projection can lie in a cluster in the Fourier transformation or vice-versa.
Such a sample then holds interesting information to base the further analysis on in the local inspection.
So, the overview of the global exploration consists of different projections of the data, various transformations, and attribution techniques to ensure an interactive exploration of many analysis spaces.


\textbf{Local Inspection --}
Local explanations consist of model decision explanations for single samples, e.g., the feature importance scores of such a sample~\cite{guidotti_survey_2018}.
For instance, attribution methods that generate local explanations are often easily applicable to many models (e.g., LIME~\cite{ribeiro_why_2016}) after training generating attributions.
Such attributions attribute a score to every input value based on the models output prediction.
Thus, such techniques are a solution to extend models with explainable components to reveal their decision making~\cite{samek_evaluating_2017}.
Explanations for images and text often show the pixels and words with heatmap techniques to highlight their importance or relevance for the prediction~\cite{guidotti_survey_2018}.
Such a technique is also possible for time series, but provides only insights into certain time points and not regions~\cite{schlegel_towards_2019}.
Thus, providing only such a heatmap requires domain knowledge to get insights into the model's decision making.

So, to support users without much domain knowledge, we enhance the local inspection with a what-if toolkit similar to the What-If tool~\cite{wexler_what_2019}.
The toolkit consists of various methods to modify the time series of a selected sample.
Further, counterfactuals of the time series can be generated to enhance the methods possible for users.
Counterfactual explanations modify a selected sample as little as possible to flip the class predicted by the classifier.
Thus, even without domain knowledge about the data, users can investigate into the models predictions using such counterfactual.
If counterfactuals do not help to dig into the model decisions, another options in the toolkit is the search for nearest neighbors based on, e.g., the euclidean distance or the models last layer activations.
Through this tool, users can explore similar samples and their features.

%% file: content/4-2-demonstrator.tex
\section{Visual Analytics Workspace} 
 
\begin{figure*}[h!t]
 \centering 
 \includegraphics[width=\linewidth,trim={0cm 7.6cm 3.3cm 0cm},clip]{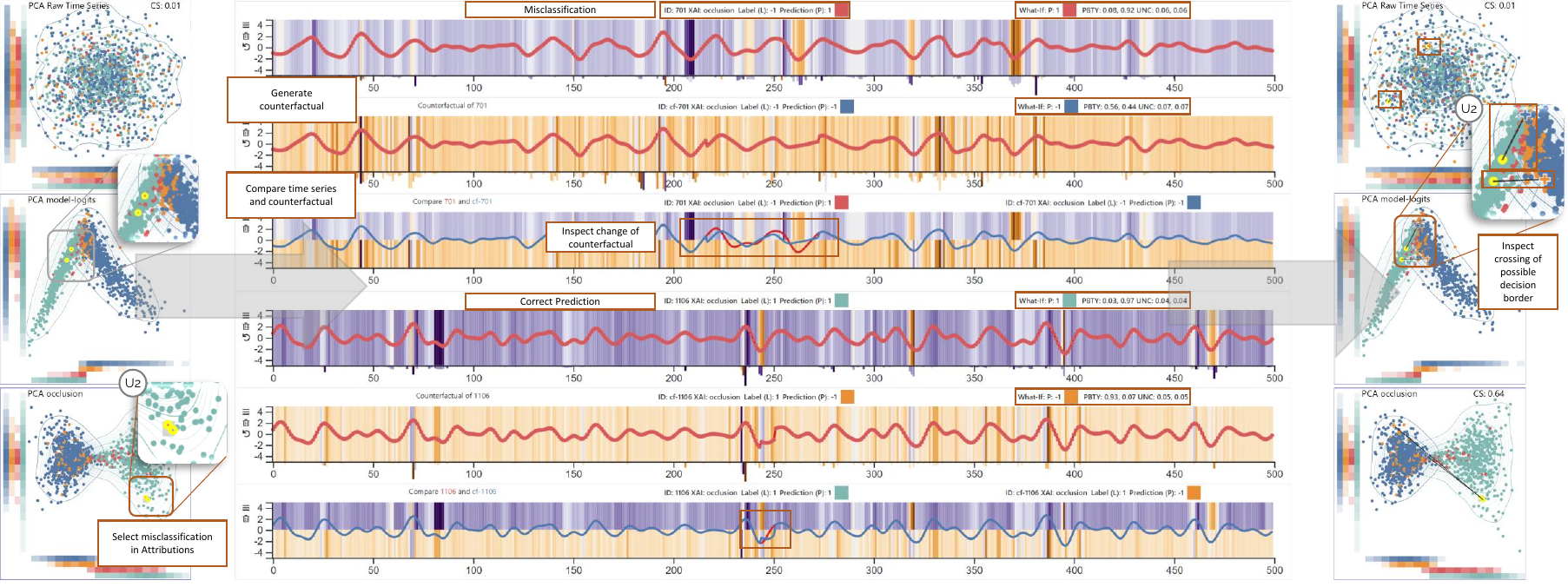}
 \vspace{-2em}
 \caption{Global to local to global transitions: selecting interesting samples in the global projections for further analysis in the local what-if. Generating counterfactual samples and compare the counterfactual changes to the original time series. Cross and line from starting time series to the counterfactual show change of position in the global projection to inspect possible decision border.}
 \label{fig:global-local-explanation}
 \vspace{-1em}
\end{figure*}


Our application\footnote{Source code online available: \newline\hspace*{8mm}{\tiny \url{https://github.com/visual-xai-for-time-series/visual-explanations-on-time-series-classification}}}, an instantiation of our proposed system and workflow, consists of four views (settings, sessions, global projections, local what-if).
The different views provide, besides their functionality, a descriptive entry point for the proposed workflow.
The settings view enables to start a new automatic phase of the workflow for a selected model and data with user set parameters.
The sessions view is the entry point after loading the application and presents the results of the automatic phase for various models to facilitate the selection of a session.
The global projections view enables the analysis of the session's global explanations to, e.g., select a misclassification.
Together with the global view, the local what-if view enables an in-depth analysis of explanations using attributions and counterfactuals.

\subsection{Settings and Sessions}
Both views are only supporting views with limited importance for the workflow.
However, these views build the backbone of the application to analyze models and data.

\textbf{Settings --}
The settings view facilitates parameter selections for the preliminary automatic phase of the workflow.
It first enables analysts to select or upload data and models.
Possible attribution methods are presented to apply to the model and the data to generate explanations.
We include DeepLIFT~\cite{shrikumar_learning_2017}, grad*input~\cite{shrikumar_not_2016}, Integrated Gradients~\cite{sundararajan_axiomatic_2017}, LIME~\cite{ribeiro_why_2016}, LRP~\cite{bach_pixel_2015}, Occlusion~\cite{zeiler_visualizing_2014}, Saliency Maps~\cite{simonyan_deep_2013}, SHAP~\cite{lundberg_unified_2017}, and Shapely Sampling~\cite{castro_polynomial_2009}.
Since time series are non-intelligible data, the settings view enables to also apply transformations on the data. 
Our selected transformations include Fourier~\cite{bracewell_fourier_1986}, Symbolic Aggregate Approximation (SAX)~\cite{lin_symbolic_2003}, Discrete Cosine transformations (DCT)~\cite{ahmed_discrete_1974}, as well as first and second-order derivatives of the time series.

The settings view further introduces parameters for the automatic evaluation methods presented by Schlegel et al.~\cite{schlegel_towards_2019}.
Notably, the evaluation techniques (point and time perturbation) can be selected to be applied to the model and the data.
Further, the thresholds for these perturbations can be adjusted to the data, e.g., the relevance threshold in a static percentage or data relative value.
Also, the sub-sequence span can be set for time evaluation techniques.
Lastly, the necessity for a check against randomization of relevance time points of the time series can be set to compare against random explanations~\cite{mittelstadt_explaining_2019}.


\noindent\textbf{Sessions -- }
The sessions view is the main entry point in which pre-computed sessions can be selected to further work with them.
These sessions retain different parameters, settings, and options for analysis in tabular format.
The columns for the table are data as well as model name, selected XAI methods, evaluation settings, and results.
The automatic results of the pre-ranking evaluation are shown as a heatmap, sorted by the performance from best to worst. 
The comparison between randomization and explanations evaluation highlights in the heatmap the usefulness of the parameters and the attribution methods on the model and data.
Through the sorting, the results are also highlighted to present the best performing method to start in the global projections.


\subsection{Global Projection View}


The global projection view visualizes the automatic phase data and results of the previously selected session in matrix form to present as much information as possible.
The first column,~\autoref{fig:global-explanation} (A), displays the raw time series projected by our selected techniques to show a first distribution of the data.
The columns after the raw data,~\autoref{fig:global-explanation} (B), consist of the selected transformations applied to the time series and enable a focus on the comparison of transformation to raw data.
The next column,~\autoref{fig:global-explanation} (C), presents the activations of the last layer before a potential softmax of the model to support users in identifying potential decision borders.
After these columns,~\autoref{fig:global-explanation} (D), the projected attribution methods are visualized and sorted by their automatically evaluated performance ranking based on the perturbation analysis.
Rows of the matrix consist of the most common projection techniques, in our case, PCA~\cite{pearson_liii._1901}, KernelPCA~\cite{schoelkopf_kernel_1997}, ISOMAP~\cite{tenenbaum_global_2000}, LLE~\cite{roweis_nonlinear_2000}, IPCA~\cite{ross_incremental_2008}, t-SNE~\cite{maaten_visualizing_2008}, LSA~\cite{halko_finding_2009}, and UMAP~\cite{mcinnes_umap_2018}.
Due to each of the various data inputs for the projection techniques having different data distributions, estimating the 
perfect fitting parameters is not trivial.
Thus, we incorporate different techniques to include various properties for the projections using default parameters to ensure some valuable projections.
Through these, we can mitigate projection artifacts as clusters shared over multiple projection techniques (e.g., manifold and linear) visualize common group properties of the samples. 

However, such an overview can get quite large with many scatter plot cells, e.g., for eight projection techniques, nine attribution techniques, and five transformations, we get more than one hundred cells.
To support users and limit the number of shown projections, we use three metrics to calculate another score to hide or show such projections.
Preferable visualizations for projections, in our case, have well-separated clusters with a high sample density.
Thus, we incorporate the Davies-Bouldin score~\cite{davies_cluster_1979}, which favors clusters far apart and less scattered.
We further want to penalize overlapping clusters and wrong-clustered points. 
Therefore, we want to give high scores to complete clusters far apart.
Thus, we introduce the euclidean distance based on the centroids of the clusters as another score, which needs to be far apart to fulfill our previous requirement.
We apply these two measures to the projections using the ground truth labels and the predictions to get four scores.
Through experiments, we weight the prediction scores higher than the ground truth labels to highlight separating projections.
The calculated \textit{cluster score} can be seen around the scatter plot color-coded and as a description at the top, see \autoref{fig:global-explanation}.
Also, projection techniques with bad scores will be hidden and other bad scores are made smaller, e.g., \autoref{fig:global-explanation}.

Every exploration can be facilitated by using colors for different sample properties focusing on a task, e.g., to identify misclassifications.
Thus, the global projections view enables three color schemes to support further analysis.
The first color schema utilizes the ground truth labels of the data to color the plot.
For up to twelve classes, a qualitative color schema is applied. 
Afterward, the color scale is exchanged to an interpolated diverging color schema due to color separability.
Such a color schema visualizes the distribution of the ground truth regarding the transformed data.
The second color applies the predictions as colors on the visualizations to enable users to analyze model boundaries.
Through such a color scale, the model predictions can be compared in other transformations to support users in understanding model and transformations correlations.
The third color schema corresponds to the confusion matrix of the data and the model.
Because the confusion matrix is a cross-product of the labels, a qualitative color scale can only be used for up to three labels.
In such a case, the confusion matrix gets overlaid by an interpolating diverging color schema starting from the top-left to the bottom-right.
Through a confusion matrix color schema, samples with distinctive properties, e.g., false positives, can be easier identified.

Through the automatic phase, the projections are in two-dimensional space to break down a lot of information towards the different transformations and need a tailored visualization.
A scatter plot visualizes each projection data 
\begin{wrapfigure}[8]{r}{2.5cm}
\includegraphics[width=\linewidth]{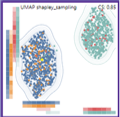}
\end{wrapfigure}
using one dot for every sample.
A contour plot extends the visibility of the clusters and how developed these are 
A dense pixel visualization on the left and on the bottom forms a distribution plot on the x- and y-axis.
The confusion matrix color schema dyes these distribution plots to enable users to, e.g., identify false positives.
Enabling filtering, selection, and comparison interactions is vital to explore global projections, as the visualization can be crowded.
Hovering or brushing over samples highlights these in the other visualizations to better compare clusters and outliers.
Further, hovering over one sample shows a tooltip of the properties of the sample, e.g., predictions, see \autoref{fig:global-explanation}.
Brushing over samples adds these to the local what-if view for an in-depth analysis.

\subsection{Local What-If Analysis}
After brushing samples in the global projections, time series visualizations for these samples are added to the local what-if view.
However, if no interesting samples are selected in the global view, we need other options for choosing compelling samples.
Thus, we include a filtering mechanism through a confusion matrix, which enables to focus the inspection to, e.g., correct predicted samples to get insights into the classification and to wrong predictions to improve the model or data
At first, the analyst can select a cell of the confusion matrix to filter for samples in the cell.
Next, the selected samples are shown as index listings to focus on specific indices.
These index listings show the standard deviation of the attributions of the sample of a selected XAI method, which defaults to the best performing.
After selecting an index, the corresponding local explanations are added to the local what-if.

One proposed visualization for the local what-if utilizes for the time series local explanation a line plot with a heatmap style background, following common literature~\cite{schlegel_time_2021}.
The local view visualizes the time series data corresponding to this technique as a line and the explanations as a background color indicating the attributions.
\begin{wrapfigure}[4]{r}{6cm}
\includegraphics[width=\linewidth,trim={7cm 0cm 0cm 0cm},clip]{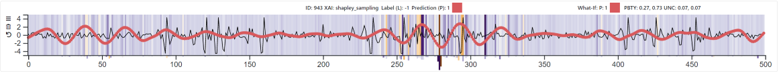}
\end{wrapfigure}
The color scale is L2 normalized on the attribution scores of the sample.
In addition to the heatmap plot of the sample, the activation maximization for a selected class label can be shown.
Our activation maximization is calculated for the classifier layer before the \textit{softmax} to generate a time series activating a selected class prediction.
A comparison between the activation maximization and the time series often already shows some overlaps or differences.

\begin{figure}[h!t]
 \centering 
 \includegraphics[width=\linewidth,trim={0cm 6cm 23cm 0cm},clip]{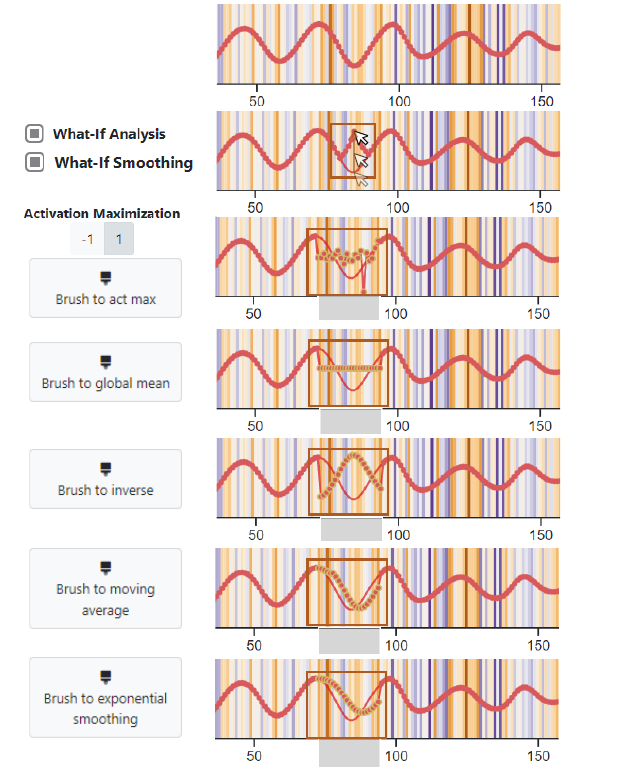}
 \vspace{-1em}
 \caption{The local what-if supports the modification of individual points with a smoothing around the dragged point, the brushing to the activation maximization, to the global mean, the local mean, the inverse, the moving average, and the exponential smoothing of the selected area.}
 \label{fig:local-what-if}
 \vspace{-1em}
\end{figure}

After being able to explore local explanations, users want to modify the sample and predict the changed version to test the model in a what-if scenario.
Thus, the view enables a what-if analysis by dragging and dropping time points to facilitate the investigation, compare~\autoref{fig:local-what-if}.
For instance, the prediction can change by altering time points with high relevance based on the attribution.
Using dropouts in the model, we further support the new prediction with the proposed uncertainty estimation by Gal and Ghahramani~\cite{gal_dropout_2016}.
We enable the smoothing of time points around possibly changed points algorithmically with a user-defined range to support users in the tedious task of time series adaption without destroying the flow.
Further, users can brush the time series and set the brushed time points to other properties.
Through interviews, we support users in setting the selected content to: a user-selected class activation maximization, the global mean, the local mean, the inverse, the moving average, and the exponential smoothing of the brushed time points as seen in~\autoref{fig:local-what-if}.
Users can use these tools to change time series in a WYSIWYG editor-like style to enable easier what-if changes.

Even with smoothing, a manual modification can get tedious. 
Thus, we support additional tools to either change or compare time series samples.
Users can search for nearest neighbors for a sample based on the Euclidean distance, the models' activations, and the attributions to get similar time series and get an explanation by example, which helps in some cases~\cite{jeyakumar_how_2020,schlegel_time_2021}.
As attributions alone are, in most cases, rather tedious and potentially misleading explanations~\cite{schlegel_time_2021}, we further include the counterfactual explanation algorithm by Delaney et al.~\cite{delaney_instance_2021}.
Delaney et al.~\cite{delaney_instance_2021} use a native guide (the nearest neighbor in the dataset) to guide the generation of a counterfactual to avoid the high dimensional optimization needed in other methods, e.g., Wachter et al.~\cite{wachter_counterfactual_2017}.
Further, two line plots can be selected and compared into a single visualization with two time series lines to directly compare the changes, differences, and similarities.
At last, newly generated time series can then be projected back into the global explanations to explore the neighborhoods around them and how these compare to their origins.
Through these interactions, users can fulfill their needs to, e.g., find decision borders.


%% file: content/5-evaluation.tex
\begin{figure*}[h!t]
 \centering 
 \includegraphics[width=\linewidth,trim={0cm 10.4cm 0cm 0cm},clip]{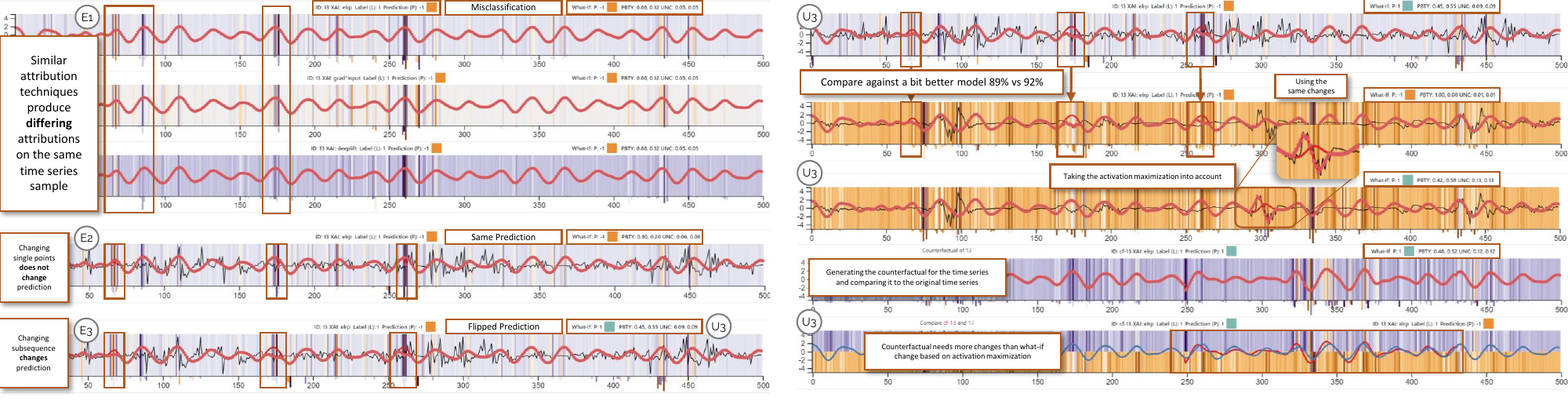}
 \vspace{-2em}
 \caption{\textbf{Left side}: First three line charts depict a diverse set of attribution techniques with different attributions for the same time series sample. Forth line chart presents the sample modifications on a few time points without a major change in the prediction. The fifth line chart demonstrates a focused modification of the time series using smoothing to achieve a prediction change. Forth and fifth include the activation maximization of the model. 
 \textbf{Right side}: First line chart presents the misclassification and the changes to flip the class. The second shows the same sample predicted with a better model and the same time point modifications as before, but the prediction does not change. Still, it is misclassified. The third line chart visualizes another modification of the time points based on the activation maximization for this model. And thus, the prediction changes. Forth and fifth show the generated counterfactual by Delaney et al.~\cite{delaney_instance_2021} with more changes needed for the flip.}
 \label{fig:local-inspection}
 \vspace{-1em}
\end{figure*}

\section{Evaluation}
\label{sec:evaluation}
To verify our approach to enable analysts to (1) explore data transformations and feature relevance, (2) identify model behavior and decision boundaries, as well as (3) reason for misclassifications, we present three use cases.
Through these cases, usage scenarios of the workflow are shown, and a particular focus on the interactions between global and local explanations is set.
At first, we discuss our experience by working closely with experts to incorporate domain knowledge into the XAI process.
Afterward, we show our selected use cases to tackle the tasks above.

\subsection{Expert Study}
We collected expert user feedback from two signal processing engineers during the development to first extract essential needs and later to tackle important analysis features.
As a baseline dataset, we used the FordA~\cite{dau_ucr_2018} dataset, as it resembles data our experts work with.
The FordA dataset is a benchmark dataset for time series classification and forms a binary anomaly detection problem.
The dataset contains 3601 training and 1320 test samples with a length of 500 time points for each sample.
These samples consist of a measurement of engine noise to classify anomalies.

Most models perform quite well on the dataset, and state-of-the-art accuracy results in $96,54$\%~\cite{dau_ucr_2018}.
Our relatively simple model achieves a test score of $89,31$\%.
We use Tensorflow~\cite{abadi_tensorflow_2015} as a deep learning library to create and train our model.
The architecture starts with three Conv1D layers with filters of (3, 6, 9), a kernel size of 3, and a stride size of 1.
We use MaxPooling with a size of 5 after each of the Conv1D layers to reduce the size of the outputs.
After these layers, we flatten the output and apply a Dense Layer with 50 neurons, dropout of 0.5 as well as ReLu activation.
Afterward, we use a softmax layer, results in our prediction.
We train our model for 100 epochs with Adam-optimizer default parameters.
We choose such an architecture as training time is quite short ($<3$ mins), and performance on test data is useful.
The architecture can be fine-tuned further, but generally solves the task.

\textbf{Exploration of data, model, and explanations --}
Fig.\ref{fig:global-local-explanation} visualizes a global representation of the FordA dataset on the left showing the raw data, the models' activation, and the attributions of Occlusion~\cite{zeiler_visualizing_2014} projections.
The global explanation presents a division of the predictions of the model into two clusters as correct (true positive) \cssclr{4e79a7} as well as wrong (false positive) \cssclr{f28e2c} predictions of anomaly build a cluster and correct (true negative) \cssclr{76b7b2} as well as wrong (false negative) \cssclr{e15759} results of non-anomaly the other. 
At first, our analysts want to get more insights into the data to find and correct possible errors.
Thus, by inspecting the global explanation and the projections of the raw time series and the transformations, they are able to analyze the predictions and data distributions in more detail,~\autoref{fig:global-explanation} (A), (B), and (C).
After examining the raw data and other transformations (Fourier and DCT), the experts suspect that the model did not learn a transformation function but some other interesting features.
By further analyzing the feature attributions of, e.g., the PCA projections,~\autoref{fig:global-explanation} (D) third row, our analysts see a diverging pattern between the predictions.
The feature attributions demonstrate a mental representation of the model applied to the data.
To their surprise, the projections show clear clusters of the model towards the predictions of the classes.
After hovering over a point in the critical region, our experts want to investigate the class anomaly of the time series sample.
By selecting the point, the corresponding time series sample is added to the local explanation, and our analysts are now able to analyze them in detail,~\autoref{fig:local-inspection}.
At first, they note that the attributions differ from each XAI technique,~\autoref{fig:local-inspection} (E1) LRP, grad*input, DeepLIFT.
After the initial thoughts, our analysts want to understand how and why the sample got its prediction.
Initially, they modify single time points with high relevance for the attribution method,~\autoref{fig:local-inspection} (E2).
However, slowly our experts start to realize that changing single points even on highly relevant points is not enough and try to modify a region with a focus on a time point with a high relevance,~\autoref{fig:local-inspection} (E3).
After adapting this selected region to be lower (more similar to the activation maximization), the prediction changes,~\autoref{fig:local-inspection} (U3) left, and our analysts suspect that time series with lower values result in the normal class.

\subsection{Use Cases}
We present, based on the previous feedback and tasks, three usage scenarios tackling (1) the exploration of data transformations and feature importance, (2) the finding of decision boundaries, as well as, (3) the reason for misclassifications.
For these use cases, we will further use the FordA dataset.
However, we extend our previous network and fit it a bit better (92,11\%) to the data for the third task (3).

\textbf{Explore Transformations and Attributions --}
\autoref{fig:global-explanation} establishes the overview with the projections on the FordA dataset with our model.
Exploring the different cells can be tedious.
However, slowly starting with the projections of the raw data supports understanding the data.
As we see in~\autoref{fig:global-explanation} (A)(U1), the raw data does not help us with further analysis as the cluster score is relatively low.
Next, transformations, in general, can help to dig into the data,~\autoref{fig:global-explanation} (B)(U1).
In our case, the data can be somewhat grouped with the Fourier transformation and a better cluster score.
However, there is still a lot of overlap in the clusters.
So, our next exploration step is the model activations, which follows a higher cluster score and more visually pleasing clusters.
In our case, we already can see in~\autoref{fig:global-explanation} (C)(U1) that there is a split between the predicted classes.
Primarily, if we focus on the confusion matrix, we see that the clusters correspond to the models' predictions.
However, we have a considerable overlap if the model is not very sure about the predictions.
The attributions,~\autoref{fig:global-explanation} (D), split the predictions even further with a similar cluster pattern to the model activations.
Depending on the attribution and projection technique, we get good separating clusters supported by high cluster scores.
Through such an analysis, users can explore their applied data a bit more and look into the representation of the model's behavior with the data.
Remarkably, the attributions enable us to gain insights and help to identify fascinating samples of, e.g., misclassifications,~\autoref{fig:global-explanation} (D)(U1), and decision borders~\autoref{fig:global-local-explanation} (U2).

\textbf{Finding Decision Boundaries --}
After exploring the data and how the model handles the data, we want to further look into the model and, e.g., find decision borders.
For such a case, we incorporate parts of the analysis of our domain experts.
In~\autoref{fig:global-local-explanation} on the left, we first look at the raw time series. 
As we have seen above, such a projection does not help much.
Next, we can look into the model activations,~\autoref{fig:global-local-explanation}, and see that we already get two clusters with a rather large overlap area.
We generate the hypothesis that a decision border divides the clusters.
To find the decision border interactively, we select a misclassification and a correct prediction to further look into the models' predictions.
In our local what-if~\autoref{fig:global-local-explanation} middle, we inspect the prediction probabilities and see that both predictions are quite confident towards their class.
Thus, we use the algorithm by Delaney et al.~\cite{delaney_instance_2021} and generate counterfactuals for these samples.

We further compare the generated counterfactual explanations to our initial time series.
In the misclassification case, the first three line plot heatmaps on top, we see that a relatively large part of the time series needs to be changed to flip the prediction, and the model is still not that confident about the prediction.
However, if we further look into the changes, we see that the modified time points are still rather similar to the initial time series.
In the correct prediction, next three line plot heatmaps, we see a different outcome.
The model has high confidence in the new misclassification with low uncertainty.
Further, inspecting the direct comparison shows only a slight change in the time series necessary to flip the prediction.
However, we need further domain knowledge about the data to analyze the changed segments of the counterfactual explanations.
So, in ~\autoref{fig:global-local-explanation} on the right (U2), we project the newly generated time series back into the global projections to explore the representation of the model a bit further and potentially verify our generated hypothesis about the decision border.
We see that the counterfactuals really cross our previous hypothesized decision border between the clusters.
Based on our previous exploration of the attributions and the better splits in them, we see that the crossing in the attributions is even larger.
The samples change their cluster membership quite drastically,~\autoref{fig:global-local-explanation} (U2).
Thus, through the easy example, we already found a decision border in the model and the attributions, while no major changes happen in the raw time series projections.
The workflow and application support such a scenario for any time series classification model and dataset.

\textbf{Reason for Misclassification --}
After we discovered a decision border at the global level, we want to dig further into the local time points to investigate misclassifications.
Especially, interesting are regions where a slight change of the time series enables a flipping of the prediction.
An optimization-based counterfactual approach like Wachter et al.~\cite{wachter_counterfactual_2017} can potentially find such minor modifications.
However, such an algorithm can potentially also degenerate the time series in a none plausible way to achieve the class change.
Thus, we enable the user to explore the local time series to reason about misclassifications.
For our focused FordA model, we can change single time points to flip the prediction in many cases as our model learned arguably by heart some flow patterns to be essential for a class.
These changes are not ideally convincing counterfactual explanations.
Thus, a what-if analysis enables users to inspect the classification more in-depth.
Our experts showed some similar, more interesting restuls in~\autoref{fig:local-inspection} (E3) / (U3).
With just a few changes which still incorporate smooth time series, the prediction can be flipped.

However, what happens if we try something similar to a more advanced model.
We add another Conv1D layer and increase the filters for each layer to [10, 50, 100, 150] to improve the accuracy score to 92,82\%.
On the right in~\autoref{fig:local-inspection}, we can see the change from the first to the second line chart.
The attribution scores differ heavily from the two models and also the activation maximization is distinguishable.
If we change the same time points like in our previous worse model, we do not see the same flip in the prediction.
Such an observation suggests that the improved model learned some other patterns for the classification.
However, in the third line chart, we modified some other parts of the time series to resemble the line chart and can observe the wanted prediction change.
Thus, we our initial assumption seems correct, and the improved model learns something different and more robust for the prediction.
Further, the forth presents the counterfactual generated with Delaney et al.~\cite{delaney_instance_2021} and the fifth the comparison between the inital time series and the counterfactual.
We see that even the counterfactual algorithm needs more time point changes for the flip of the prediction.

%% file: content/6-discussion.tex
\section{Discussion}
\label{sec:discussion}

Our workflow for visual explanations with attributions and counterfactuals introduces seamless interactions between global and local explanations by incorporating local attributions into global projections.
To generate such global explanations, we transform local attributions to the global level by using a projection over data applied to the model.
By supporting interactions between these two levels of explanations, users can get a general overview of the models' representation and then analyze in-depth single samples and single model decisions.
Through diverse algorithms, visualization, and interaction techniques, we present an application incorporating the workflow and demonstrate the applicability of time series classification, giving users a low obstacle to exploring their models.
Using the application, our presented use cases demonstrate examples of possible tasks that can be tackled using the workflow to facilitate the analysis of models and datasets.

\textbf{Lessons Learned --}
Although demonstrating the applicability of our approach, there are some lessons we consider after developing the proposed workflow and application.
While we knew that attribution methods work on time series models and that these models learn the time component of time series, the attributions show, in most cases, only single time points as most important for the classification~\cite{schlegel_time_2021}.
We thought that the attributions would present essential regions for the classification.
Such regions facilitate the analysis of the samples by identifying possible data errors.

\textbf{Limitations --}
Our proposed workflow enables a seamless transition between global and local explanations but holds some limitations regarding the incorporated methods.
One of the most prominent limitations is the dependence on high-quality local attribution methods and their applicability to the model and the data.
We introduce a way to overcome this limitation through our automatic evaluation of the method.
However, the evaluation metrics are imperfect and depend on parameters.
Our current approach focuses heavily on time series inspection and analysis by evaluating only this data type in detail.
Nevertheless, the workflow can be applied to any data type and application with working attribution methods.

\textbf{Research Opportunities --}
Based on our identified limitations, we present research opportunities to enhance our workflow to improve generalizability.
To handle the most critical point, the evaluation of attribution methods, we argue for improved techniques to incorporate data-specific challenges, such as better zero reference points, similar to Schlegel et al.~\cite{schlegel_towards_2020}.
Further, the identification of thresholds can be automated by implementing heuristics to compare the results over the selected XAI methods and adjust the thresholds until a certain lower bound is found.
Through such an automated analysis, the inspection of the global and local explanations can be facilitated by having more descriptive scores. 

Another option for the global projections demonstrates path projections~\cite{ali_timecluster_2019}.
These projections can be used on the time series and attribution data, as these also show time importance.
Such an extension to the global explanations supports the analyst in improving their understanding of the model and especially the data.

Further, as the workflow applies to every data type with working attribution methods, support for these data types in the local inspection has to be explored as, e.g., even the implemented line chart heatmaps are not perfect~\cite{schlegel_time_2021}.
Mainly, improved heatmap concepts~\cite{schlegel_time_2021} and what-if interactions facilitate an analysis of the local level to project back to the global explanation.
For instance, decision borders in image classification can be found through an improved method in such a case, which enables a debugging of computer vision models.
Such an analysis potentially enables to identify borders between real data and adversarial images in the global explanation and steers model developers into improving datasets and models.


%% file: content/7-conclusion.tex
\section{Conclusion}
In this paper, we presented visual explanations with attributions and counterfactuals, a visual analytics workflow, and an application to explore and explain complex time series classifiers, e.g., deep learning models.
The workflow presents an automatic phase to apply and evaluate possible XAI techniques while later letting users to inspect global and local explanations in a loop to enrich both explanation levels mutually.
We further present an instantiation that incorporates the workflow with implementation variants.
We show the applicability of the workflow and the application using an expert study and use cases on various time series classification datasets and models.
The underlying workflow is not limited to time series classification and can be extended to work with any data type and nearly any model.